\documentclass{appolb}
\usepackage{epsfig}

\begin{document}
\title{$e^+e^- \to \pi^+\pi^- e^+e^-$ : a potential background \\
 for $\sigma (e^+e^- \to \pi^+\pi^-)$ measurement \\
 via radiative return method
\thanks{Presented by E. Nowak 
 at XXVII International Conference of Theoretical Physics,
 `Matter To The Deepest', Ustro{\'n}, 15-21 September 2003, Poland.
 Work supported in part by
 EC 5-th Framework Program under contract 
  HPRN-CT-2002-00311 (EURIDICE network) and
  Polish State Committee for Scientific Research
  (KBN) under contract 2 P03B 017 24.
}}
\author{Henryk Czy\.z and El\.zbieta Nowak
\address{Institute of Physics, University of Silesia,
PL-40007 Katowice, Poland.}
}
\maketitle 
\vspace{-5 cm}
\hfill{\bf TTP03-30}
\vspace{+5 cm}
\begin{abstract}
 A Monte Carlo generator (EKHARA) has been constructed to simulate the 
 reaction $e^+e^- \to \pi^+\pi^- e^+e^-$ 
 based on initial and final state emission of a $e^+e^-$ pair
 from $e^+e^-\to\pi^+\pi^-$ production diagram.
 A detailed study of the process, as a potential background
 for $\sigma (e^+e^- \to \pi^+\pi^-)$ measurement 
 via radiative return method, is presented for
 $\Phi$- and $B$- factory energies.
\end{abstract}
\PACS{13.40.Ks,13.66.Bc}
  
\section{Introduction}

 The radiative return method \cite{Binner:1999bt} (look \cite{cg:ustron}
 for a short introduction) is a powerful tool in the 
 measurement of $\sigma (e^+e^- \to \  {\mathrm{hadrons}})$, crucial
 for predictions of the hadronic contributions to $a_\mu$, the anomalous 
 magnetic moment of the muon, and to the running of the electromagnetic
 coupling from its value at low energy up to $M_Z$ (for recent reviews look
 \cite{Davier:2003,Nyffeler:2003,Jegetc.}).
  Due to a complicated experimental setup,
 the use of Monte Carlo (MC) event generators 
 \cite{Binner:1999bt,Czyz:2000wh,Rodrigo:2001kf,Czyz:2002np,Czyz:PH03},
 which includes various radiative corrections \cite{Rodrigo:2001jr,Kuhn:2002xg}
 is indispensable. Some more extensive analysis of that subject can be found
 also in \cite{Kuhn:2001,Rodrigo:2001cc,Rodrigo:2002hk}.
 The most important hadronic mode, i.e. $\pi^+\pi^-$, is currently measured 
 by KLOE \cite{Achim:radcor02,KLOE:2003,Juliet,SdiFalco:Ustron}, and BaBar
 \cite{Blinov} by means of radiative return method. This measurement
  can suffer from
 a background from the process $e^+e^- \to \pi^+\pi^- e^+e^-$,
 as suggested in \cite{Hoefer:2001mx}, for at least
 two reasons: 1. At present KLOE measures only pions (+ missing momenta)
  in the final state and for that particular measurement there is no
 difference between photon(s) and pair production. 2. The $e^+e^-$
 pair can escape detection, being lost outside a detector,
 e.g. in the beam pipe, 
 or having energy below a detection threshold.
 Again a Monte Carlo study is
 unavoidable, if one likes to know the actual value of the pair production 
 contribution in a given experimental setup, as the analytical, completely
 inclusive, calculations might lead to misleading results.
 
\section{Monte Carlo program EKHARA and its tests}

\begin{figure}[ht]
\begin{center}
\epsfig{file=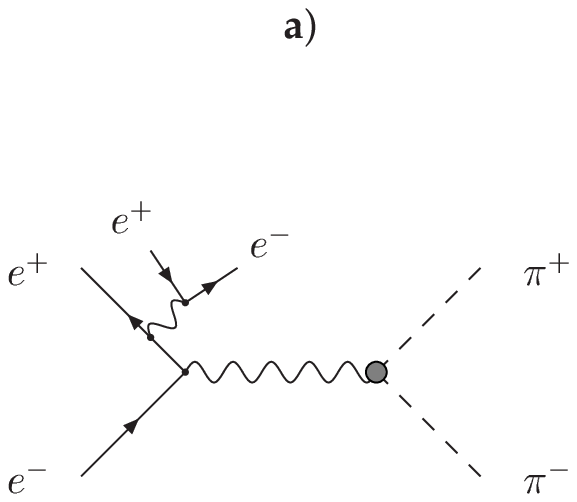,width=3.3cm,height=2.9cm}
\hskip+0.5cm
\epsfig{file=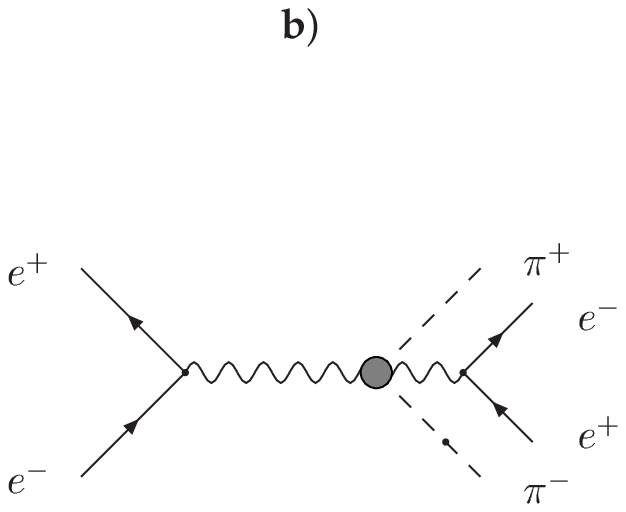,width=3.2cm,height=2.9cm}
\hskip+0.3cm
\epsfig{file=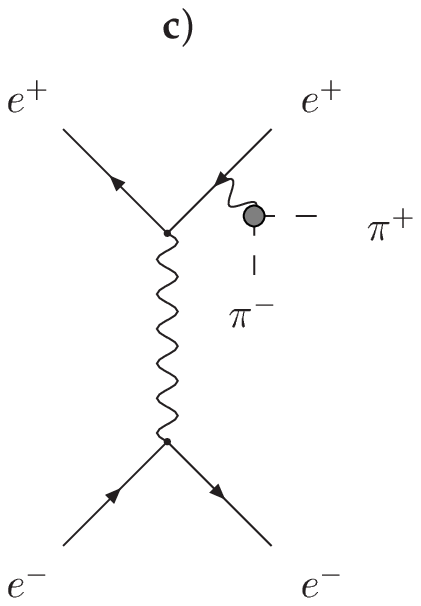,width=2.2cm,height=3cm}
\hskip+0.5cm
\epsfig{file=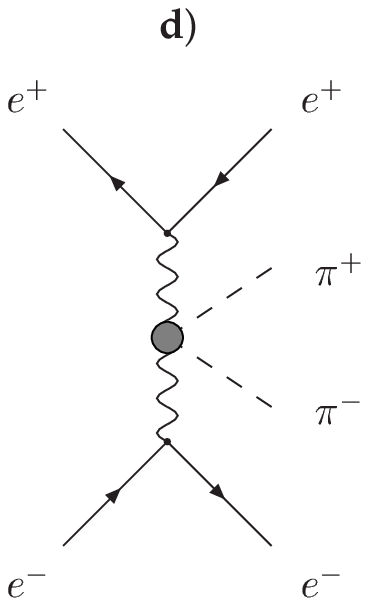,width=2cm,height=3cm}
\end{center}
\caption{Diagrams contributing to the process 
\(e^+(p_1)e^-(p_2) \to \pi^+(\pi_1)\pi^-(\pi_2)e^+(q_1)e^-(q_2)\):
 initial state electron pair emission (a),
 final state electron pair emission (b),
 pion pair emission from t--channel Bhabha process  (c) and
 $\gamma^*\gamma^*$ pion pair production (d). 
}
\label{f00}
\end{figure}

In Fig.\ref{f00} different types of diagrams
  contributing to process $ e^+e^- \to \pi^+\pi^-e^+e^- $ 
 are shown schematically.
 In the present version of the Monte Carlo program
 we include only two gauge invariant sets of diagrams from
 Fig.\ref{f00}a and \ref{f00}b. 
 The former represents initial state radiation (ISR),
 and the latter final state radiation (FSR), of an $e^+e^- $ pair 
 from  $ e^+e^- \to \pi^+\pi^-$ production diagram.
 We use scalar QED (sQED) to model FSR $e^+e^- $ pair emission
 and $\rho$ dominance model for $\gamma^*(\rho^*)\pi\pi$ coupling
 (see \cite{Binner:1999bt} for details).
 The diagrams from  Fig.\ref{f00}c, representing 
 pion pair emission from t--channel Bhabha process, together with
 s-channel Bhabha pion pair emission (not shown in Fig.\ref{f00}),
 will be included in the new version \cite{CN_new}
 of the presented generator, completing the discussion of this paper. 
 The contribution from $\gamma^*\gamma^*$ pion pair production process
 (Fig.\ref{f00}d)
 is negligible for DA$\Phi$NE energy \cite{Juliet}, and as its interference
 with other diagrams does not contribute to the cross section 
 integrated over charge symmetric cuts, these contributions are not
 relevant, at least for $\Phi$--factories.

 For parametrisation of the phase space we use the following variables:  
$Q^2=(\pi_1+\pi_2)^2$--invariant mass of $\pi^+\pi^-$ system, 
$k_1^2=(q_1+q_2)^2$--invariant mass of $e^+e^-$ system, polar and azimuthal 
 angles of $\vec{Q}$ momentum, 
defined in the centre-of-mass (CM) frame of initial $e^+e^-$ pair, 
 polar and azimuthal angles of $\vec\pi_1$ momentum, 
 defined in $Q$- rest frame and 
 polar and azimuthal angles of $\vec q_1$ momentum,
  defined in $k_1$- rest frame. All four vectors are boosted 
 into the initial $e^+e^-$ CM frame after being generated and all
 necessary cuts can be applied at this stage of the generation.
 Multi--channel variance reduction method was used to improve efficiency of the
 generator and all details
 will be given in a separate publication \cite{CN_new}.

 We have performed a number of 'standard' tests
  to ensure that the written FORTRAN code is correct.  
 Gauge invariance of the sum of the amplitudes was checked analytically
 separately for set of diagrams from Fig.\ref{f00}a and \ref{f00}b.
 We use helicity amplitudes in EKHARA to calculate square
 of the matrix element describing the  $ e^+e^- \to \pi^+\pi^-e^+e^- $
 process, but as a cross check, we have used also the standard trace
 technique to calculate
 the  square  of the matrix element, summed over polarisations
 of initial and final leptons.
 Both results were compared numerically scanning 
 the physical phase space, and the biggest relative difference
 between the two results found was  at the level of $10^{-9}$.
 It was necessary to use quadrupole precision of the real numbers
 for the trace technique result, as one can observe severe
 cancellations between various terms.
 The phase space volume calculated 
 by Monte Carlo program was cross checked
 with the Gauss integration and
 the relative difference at the level of $10^{-5}$ was well within
 the errors of the MC result, which were of the same order.

\begin{figure}[ht]
\begin{center}
\hskip-0.2cm
\epsfig{file=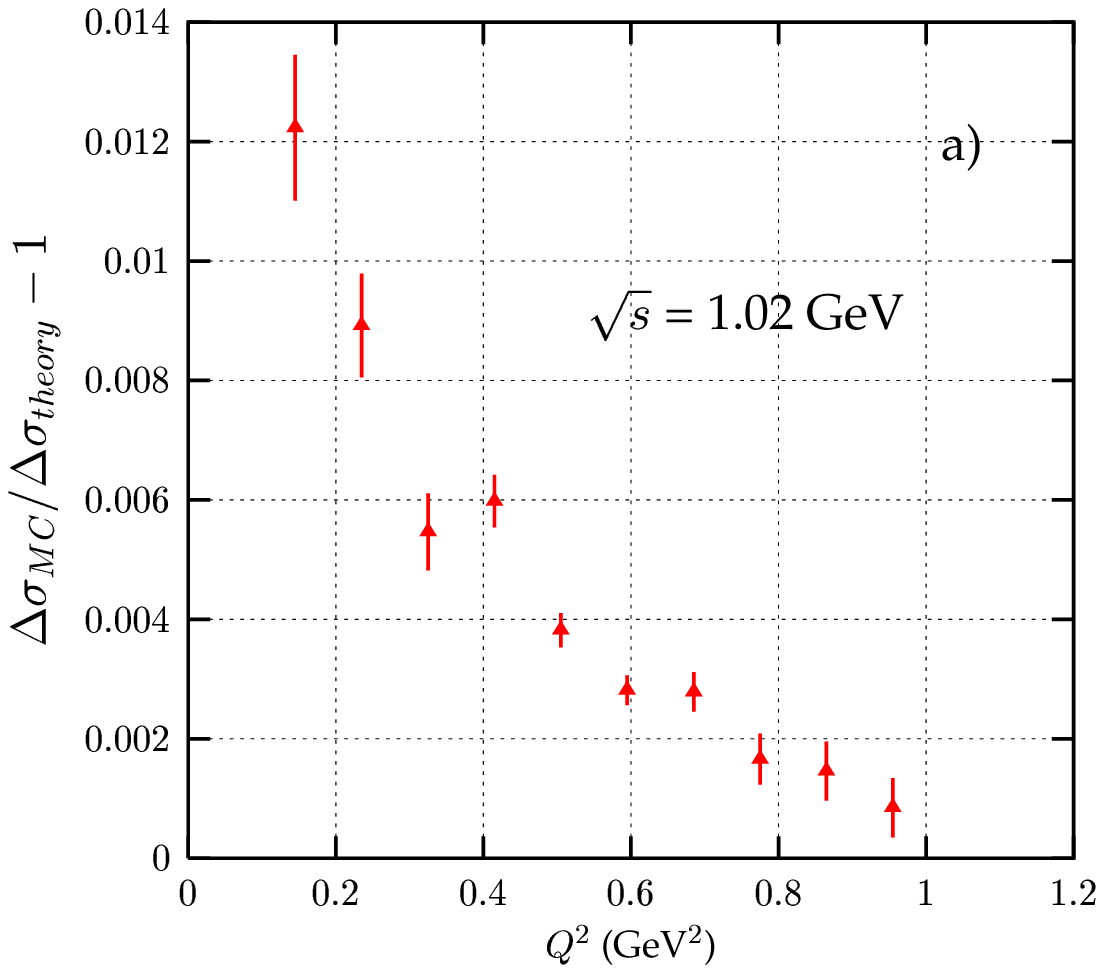,width=6.4cm,height=5.8cm}
\hskip-0.2cm
\epsfig{file=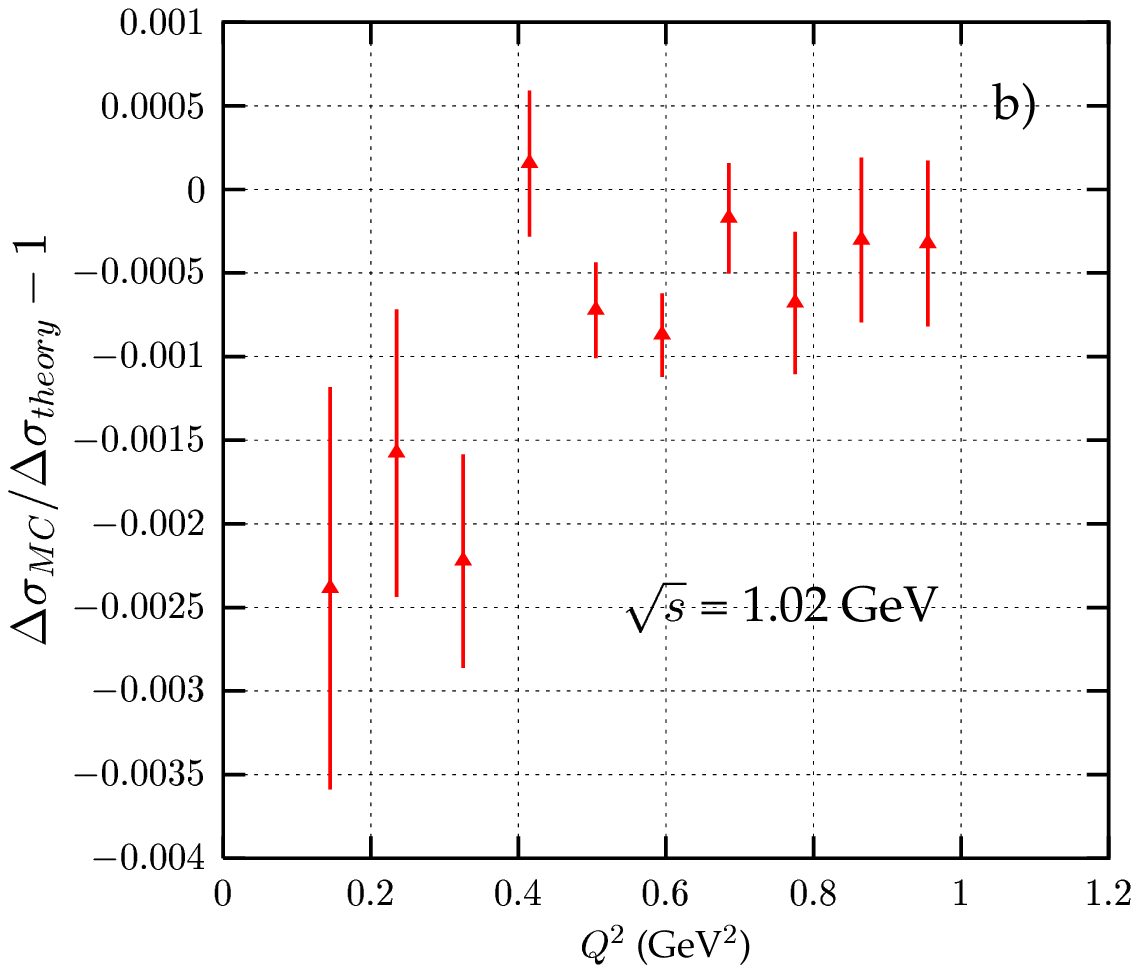,width=6.4cm,height=5.8cm}
\end{center}
\caption{EKHARA results compared with analytical results of \cite{KKKS88}
 (see text for details). The errors come from MC integration.
}
\label{f0}
\end{figure}

 Inclusive analytical formulae from \cite{KKKS88} provide additional,
 nontrivial tests of the implementation of the contributions from 
 Fig.\ref{f00}a.
 Formula (23) from \cite{KKKS88} provides $Q^2$ differential cross section
 (other variables are integrated over the whole allowed range) valid for large
 $Q^2$. In Fig.\ref{f0}a, we compare 
 the values of the integrals, over 10 equally spaced
 intervals of $Q^2$, obtained by means of MC program and one-dimensional
 Gauss integration of the above mentioned analytical formula. The 
  Gauss routine, which we use, guarantees precision of 12 significant digits.
 One observes a relatively good agreement for values of $Q^2 \sim s$
 and worse for $Q^2$ nearby $\pi^+\pi^-$ production threshold, as expected
 from the applicability of the analytical formula.
 EKHARA results agree much better (see Fig.\ref{f0}b) with known analytically
 doubly differential cross section $\frac{d\sigma}{dQ^2 dk_1^2}$
  \cite{BFK,KKKS88}, integrated over the whole allowed range
 of $k_1^2$ and 10 equally spaced intervals of $Q^2$. The exact analytical
 result was integrated numerically, using recursively one-dimensional 
 8-point Gauss procedure and dividing the region of integration
 into pieces small enough to guarantee the overall accuracy of 
 10 significant digits. From Fig.\ref{f0}b it is clear that a technical
 precision of EKHARA of the order of 0.1\% was achieved.

\section{Monte Carlo Data Analysis}

\begin{figure}[ht]
\begin{center}
\hskip-0.2cm
\epsfig{file=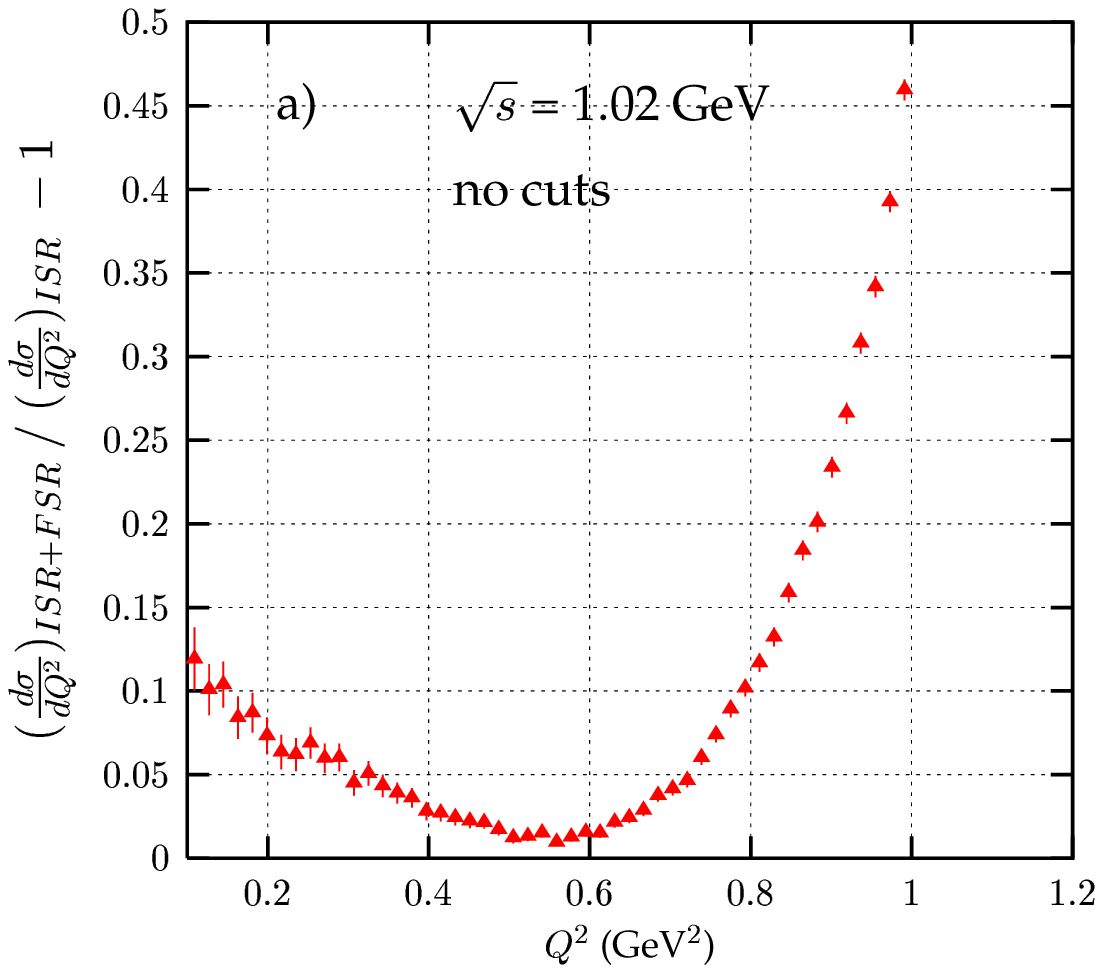,width=6.4cm,height=5.8cm}
\hskip-0.2cm
\epsfig{file=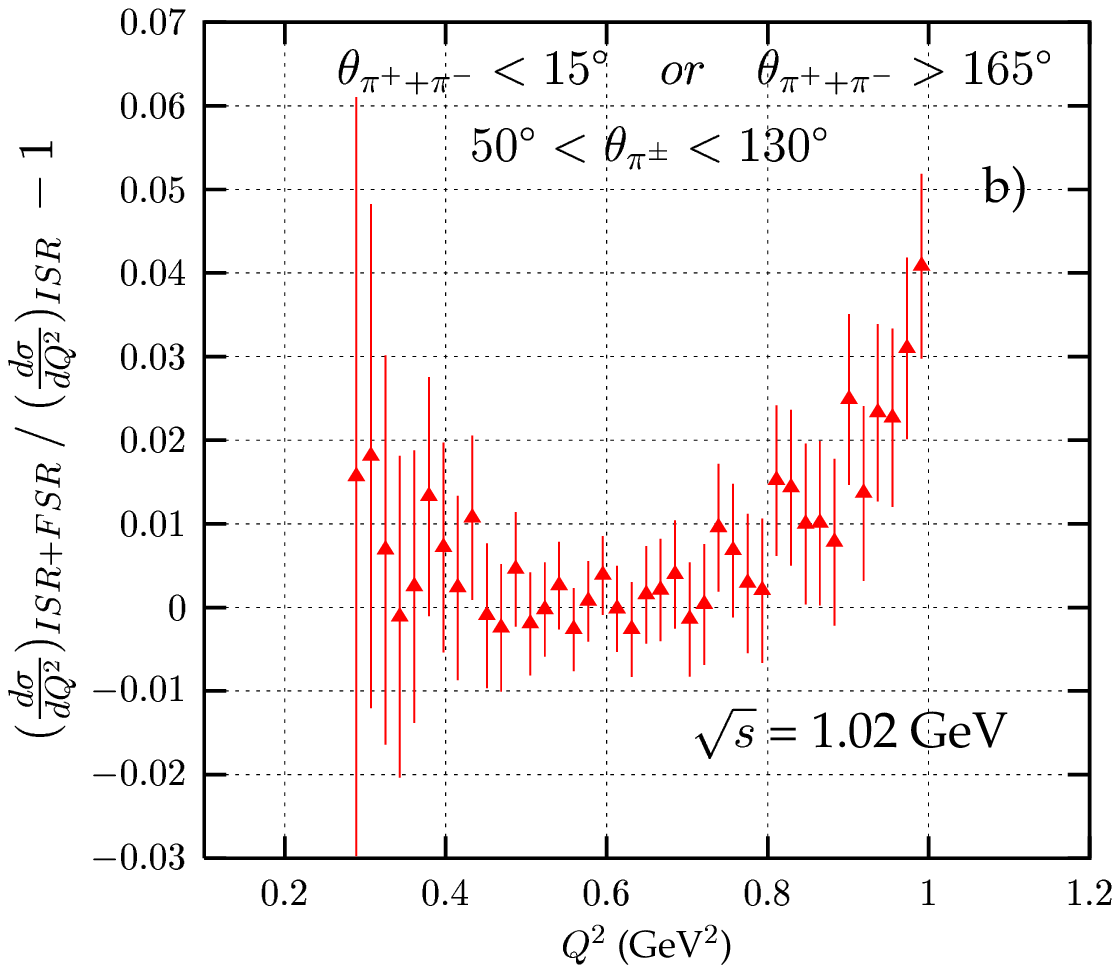,width=6.4cm,height=5.8cm}
\end{center}
\caption{Comparisons of ISR and FSR contributions
  to $e^+e^- \to \pi^+\pi^- e^+e^-$ cross section at DA$\Phi$NE energy.
}
\label{f1}
\end{figure}

 For ISR of $e^+e^-$ pair ( Fig.\ref{f00}a),
 a factorisation similar to photon emission
 holds \cite{KKKS88} and adding ISR pair production to ISR photon
 production results just in a change of 
 the radiator function, thus radiative return method
 still can be used \cite{cg:ustron}.
 On the other hand,
 FSR of  $e^+e^-$ pair ( Fig.\ref{f00}b) is model dependent, the same
 way as it is the emission of a real photon, and 
 the question of its relative, to ISR, contribution to the cross section
 is as important as for the photon emission.
 One can observe, that $e^+e^-$ pair emission resembles a lot photon emission,
 with big contributions of FSR for inclusive configurations (Fig.\ref{f1}a)
 of a $\Phi$--factory,
 which can be easily reduced, by suitable cuts, to a negligible level
 (Fig.\ref{f1}b). Moreover, the cuts which reduce photon FSR
 reduces as well  the $e^+e^-$ pair FSR.
 In addition, analogously to photon 
 FSR,  $e^+e^-$ pair FSR is completely negligible at $B$--factories. 

\begin{figure}[ht]
\begin{center}
\hskip-0.2cm
\epsfig{file=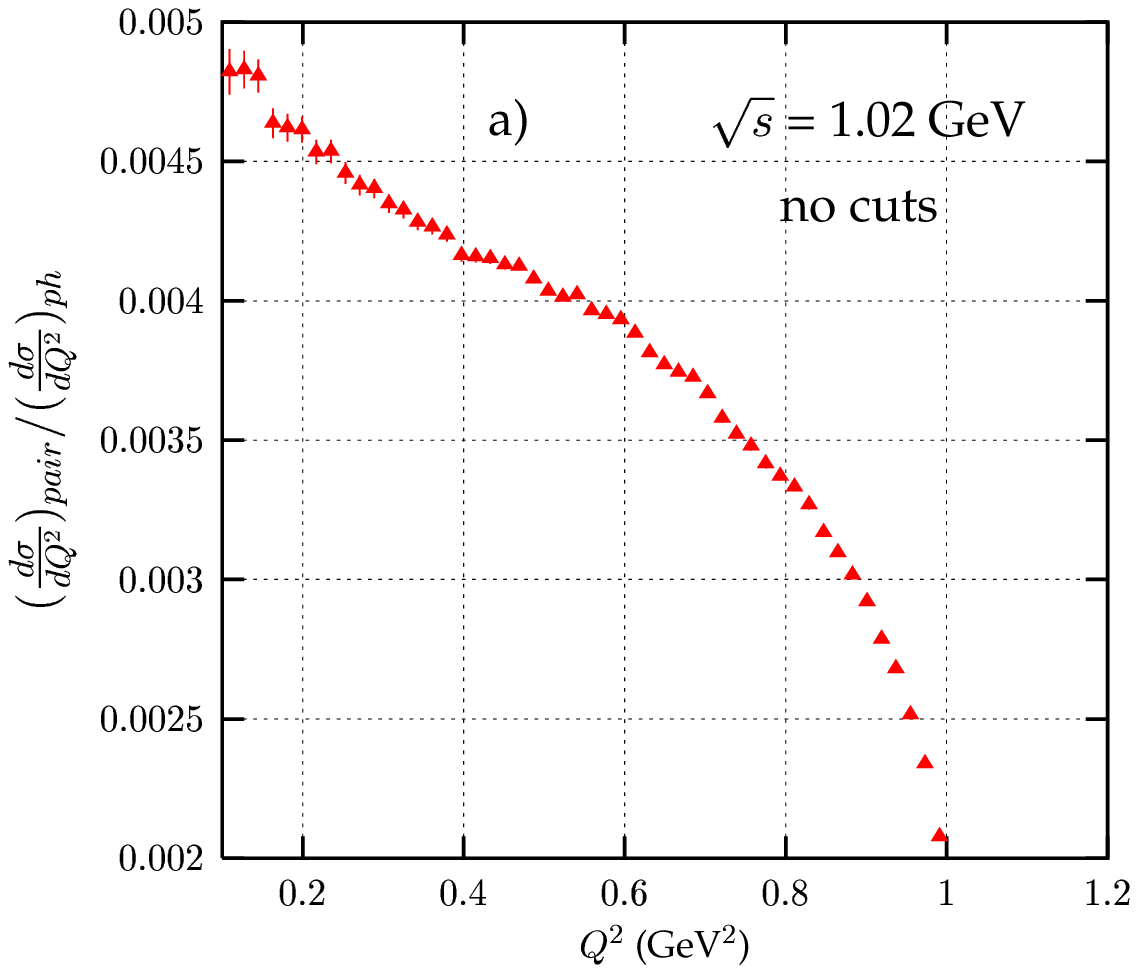,width=6.4cm,height=5.8cm}
\hskip-0.2cm
\epsfig{file=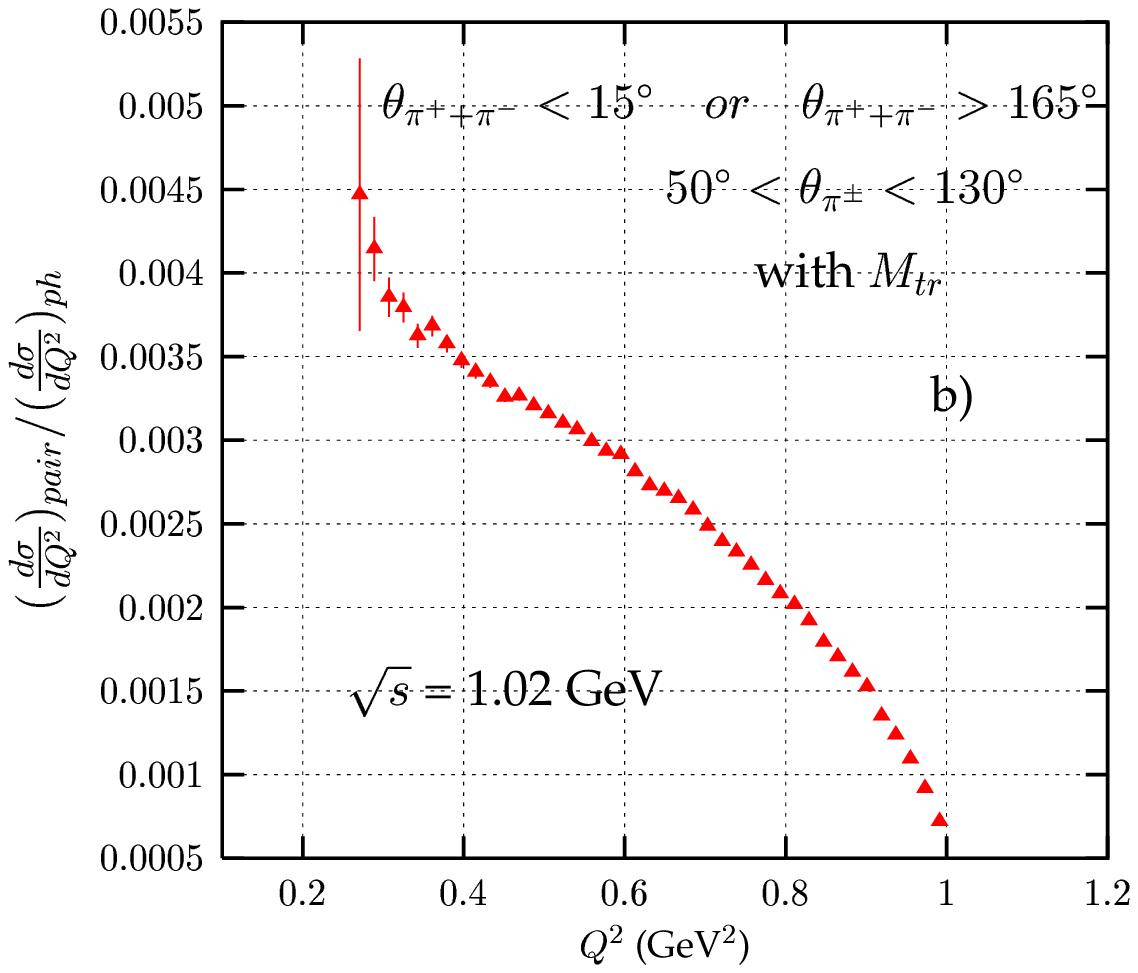,width=6.4cm,height=5.8cm}
\end{center}
\caption{ Ratio of differential cross sections 
of  the  process $ e^+e^- \to \pi^+\pi^-e^+e^- $ (pair)
 and $ e^+e^-\to \pi^+\pi^- + {\mathrm{photon(s)}} $ (ph).
}
\label{f2}
\end{figure}

\begin{figure}[ht]
\begin{center}
\hskip-0.2cm
\epsfig{file=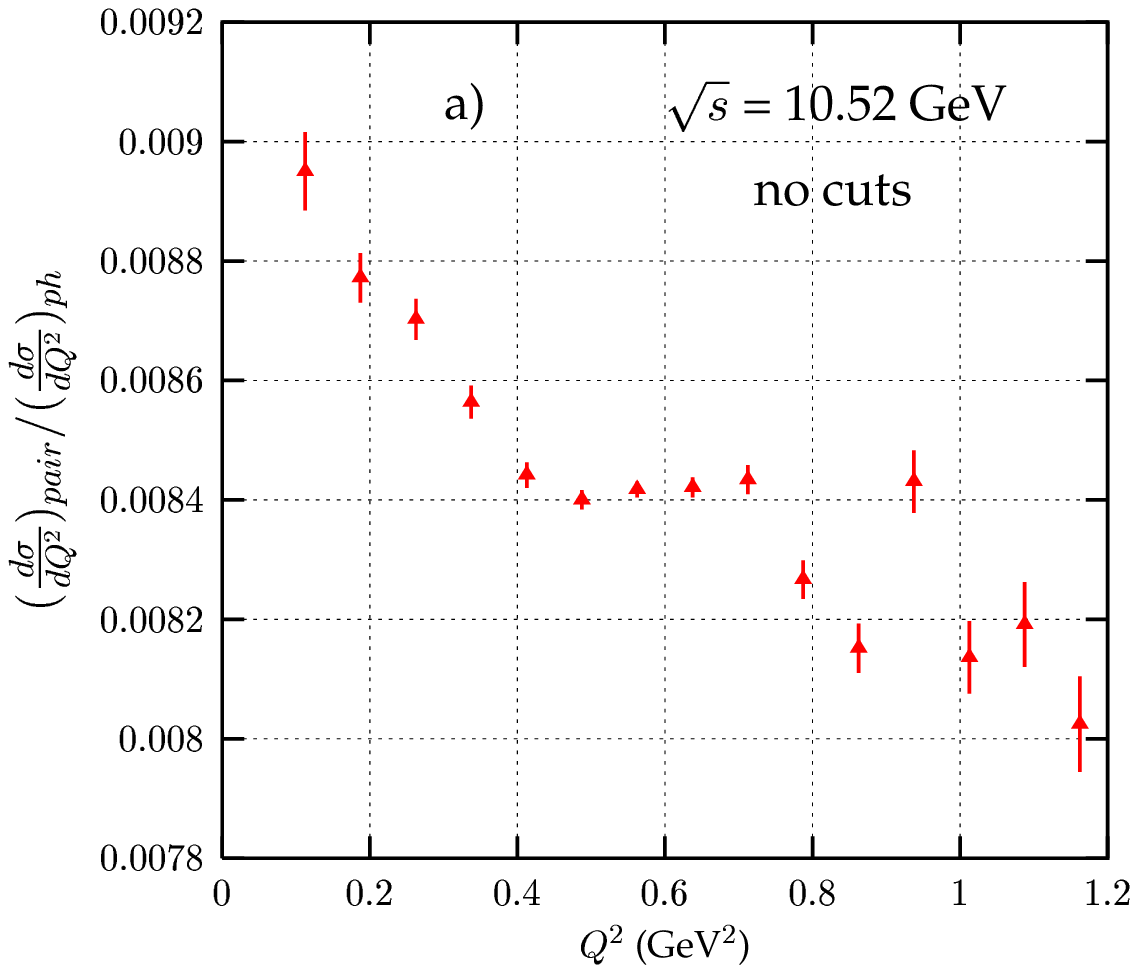,width=6.4cm,height=5.8cm}
\hskip-0.2cm
\epsfig{file=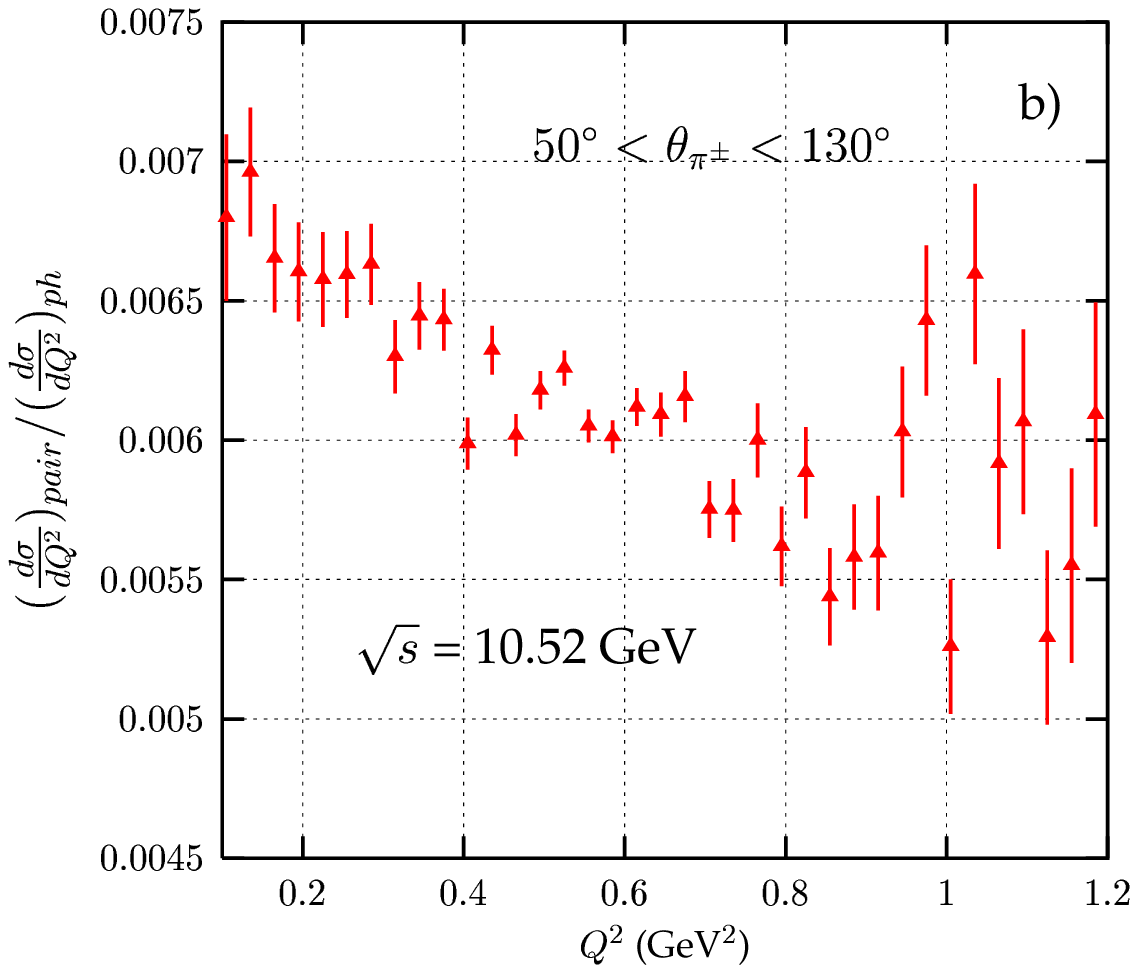,width=6.4cm,height=5.8cm}
\end{center}
\vskip-0.4cm
\caption{Ratio of differential cross sections 
of  the  process $ e^+e^- \to \pi^+\pi^-e^+e^- $ (pair)
 and $ e^+e^-\to \pi^+\pi^- + {\mathrm{photon(s)}} $ (ph).
}
\label{f4}
\end{figure}

 The most relevant information, how big is the contribution
 of  the  process $ e^+e^- \to \pi^+\pi^-e^+e^- $ in comparison to
 the main process used in the radiative return method, mainly
 $ e^+e^-\to \pi^+\pi^- + {\mathrm{photon(s)}} $,  is presented in 
 Fig.\ref{f2} for DA$\Phi$NE energy, both without any cuts (Fig.\ref{f2}a),
 and with cuts resembling KLOE event selection 
 \cite{KLOE:2003,SdiFalco:Ustron} (Fig.\ref{f2}b). The results of
 $ e^+e^-\to \pi^+\pi^- + {\mathrm{photon(s)}} $ cross section were
 obtained using PHOKHARA 3.0 MC generator \cite{Czyz:PH03} and in
 the following, whenever we refer 
 to $ e^+e^-\to \pi^+\pi^- + {\mathrm{photon(s)}} $ cross section
 we mean cross section obtained using PHOKHARA 3.0. As one can see
 from Fig.\ref{f2}, the contribution
 of the $e^+e^-$ pair production is below 0.5\%, independently on the cuts.
  It is $Q^2$ dependant, being big for
 low $Q^2$ values.
 Even if it is small,
  this 0.5\% contribution can become
 important, when aiming at the precision below, or of the order of 1\%, for the
  $ e^+e^-\to \pi^+\pi^-$ cross section measurement.
 At $B$--factories, the relative contribution of the pair production
 might be as big as 0.9\% (Fig.\ref{f4}a) and it is again $Q^2$ and
 cut dependent (Fig.\ref{f4}b).

\begin{figure}[hb]
\begin{center}
\hskip-0.2cm
\epsfig{file=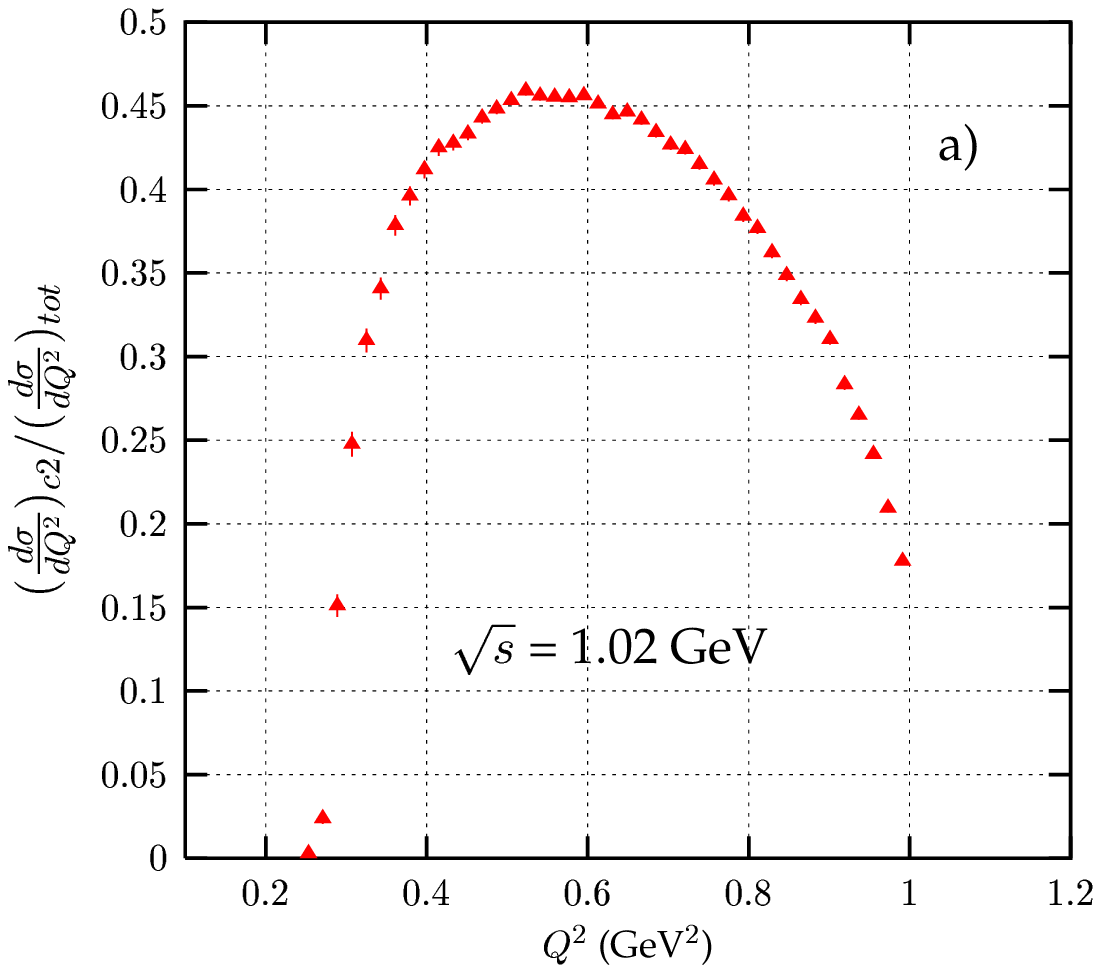,width=6.4cm,height=5.8cm}
\hskip-0.2cm
\epsfig{file=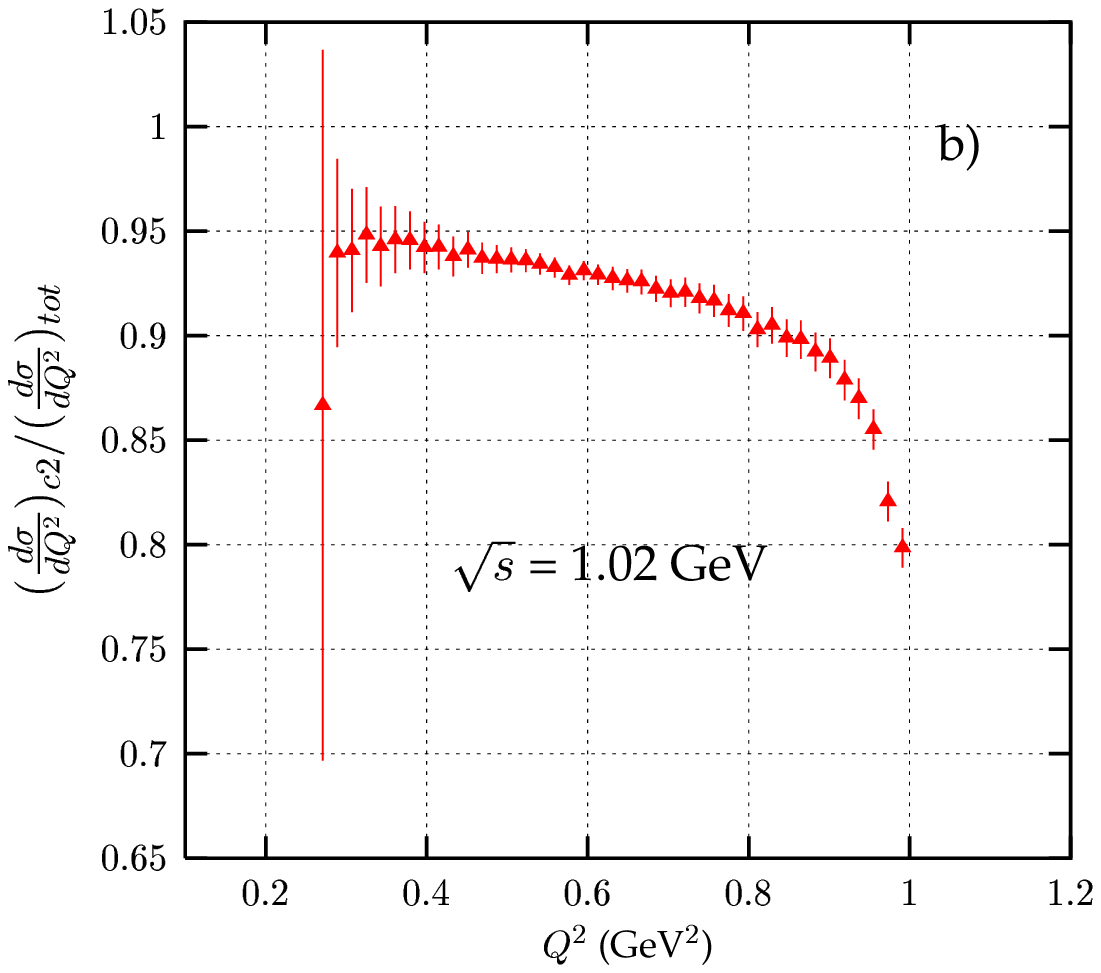,width=6.4cm,height=5.8cm}
\end{center}
\vskip-0.4cm
\caption{Non-reducible pair production background 
 at DA$\Phi$NE energy : (a) no cuts in $ ( \frac{d\sigma}{dQ^2})_{tot}$;
 (b) for both $ (\frac{d\sigma}{dQ^2})_{c2} $ and 
  $ ( \frac{d\sigma}{dQ^2})_{tot}$ the following cuts are imposed:
   $  50^\circ < \theta_{\pi^{\pm}} < 130^\circ $,
  $ \theta_{\pi^+ + \pi^-} < 15^\circ $ 
  or $ \theta_{\pi^+ + \pi^-} > 165^\circ $ and $M_{tr}$ cut.
 For (a) and (b) c2 denotes additional cuts:
 $ (\theta_{e^+} < 20^\circ $ or $  \theta_{e^+} > 160^\circ) $
and $ (  \theta_{e^-} < 20^\circ $ or $  \theta_{e^-} > 160^\circ)  $.
}
\label{f3}
\end{figure}

  As stated already, ISR of electron pairs can be treated in a similar way as 
 ISR of photons resulting in the change of radiator function in
 the radiative return method. However, one can alternatively try to treat it
 as a background to the process with photon(s) emission.
  In this case, the most natural way of reducing that
 background is to veto the electron (positron) in the final state.
 In Fig.\ref{f3} we show an example of such a procedure performed for
 $\Phi$--factory energy. We assume here that an electron or positron
 can be seen, and the event rejected, if its angle with respect to 
 the beam axis is bigger then 20$^\circ$. Fig.\ref{f3}a shows that
 up to 50\% of the events pass the rejection procedure, when no other
 cuts are applied. However, in the case of KLOE event selection, which requires
 that the $e^+e^-$ pair is emitted along the beam axis, one 
 rejects only a small fraction of these
  events (Fig.\ref{f3}b).

\begin{figure}[ht]
\begin{center}
\hskip-0.2cm
\epsfig{file=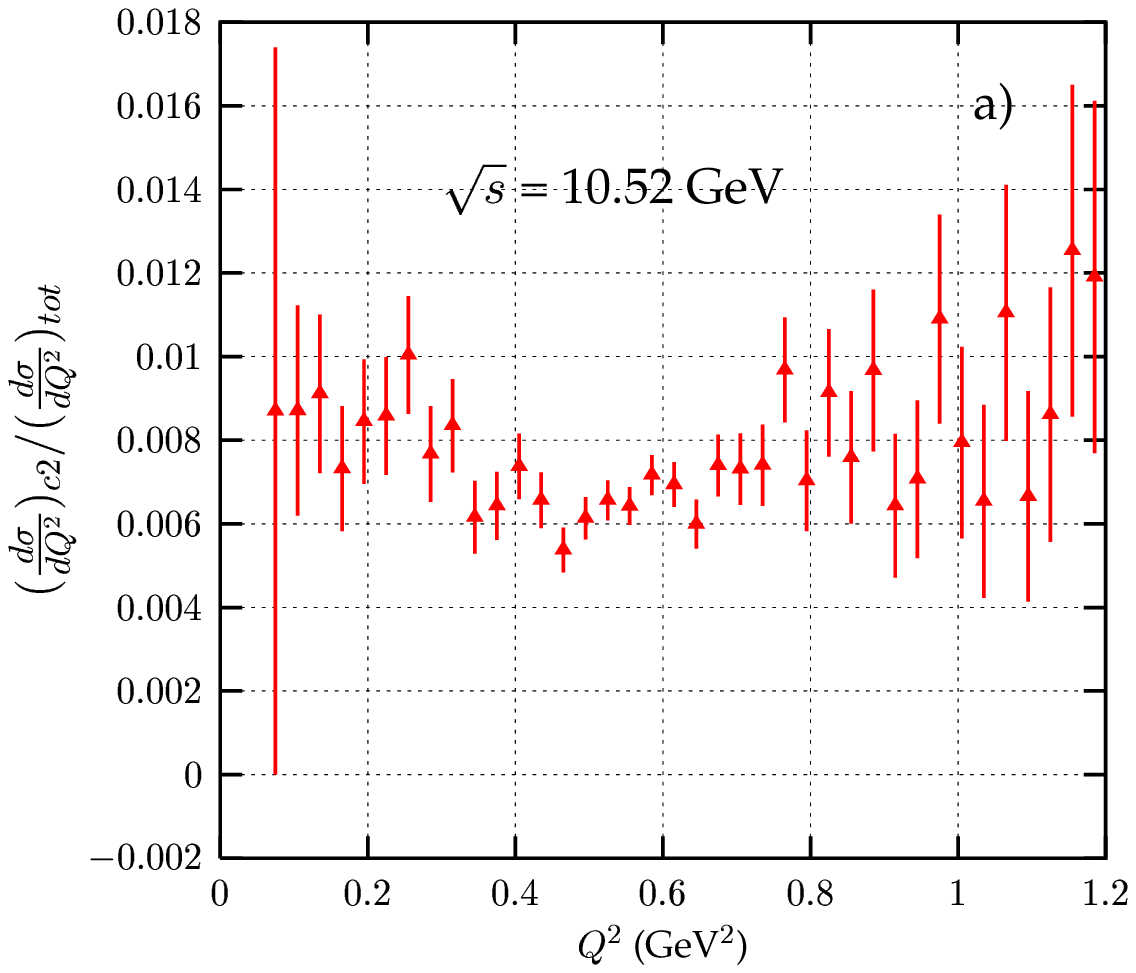,width=6.4cm,height=5.8cm}
\hskip-0.2cm
\epsfig{file=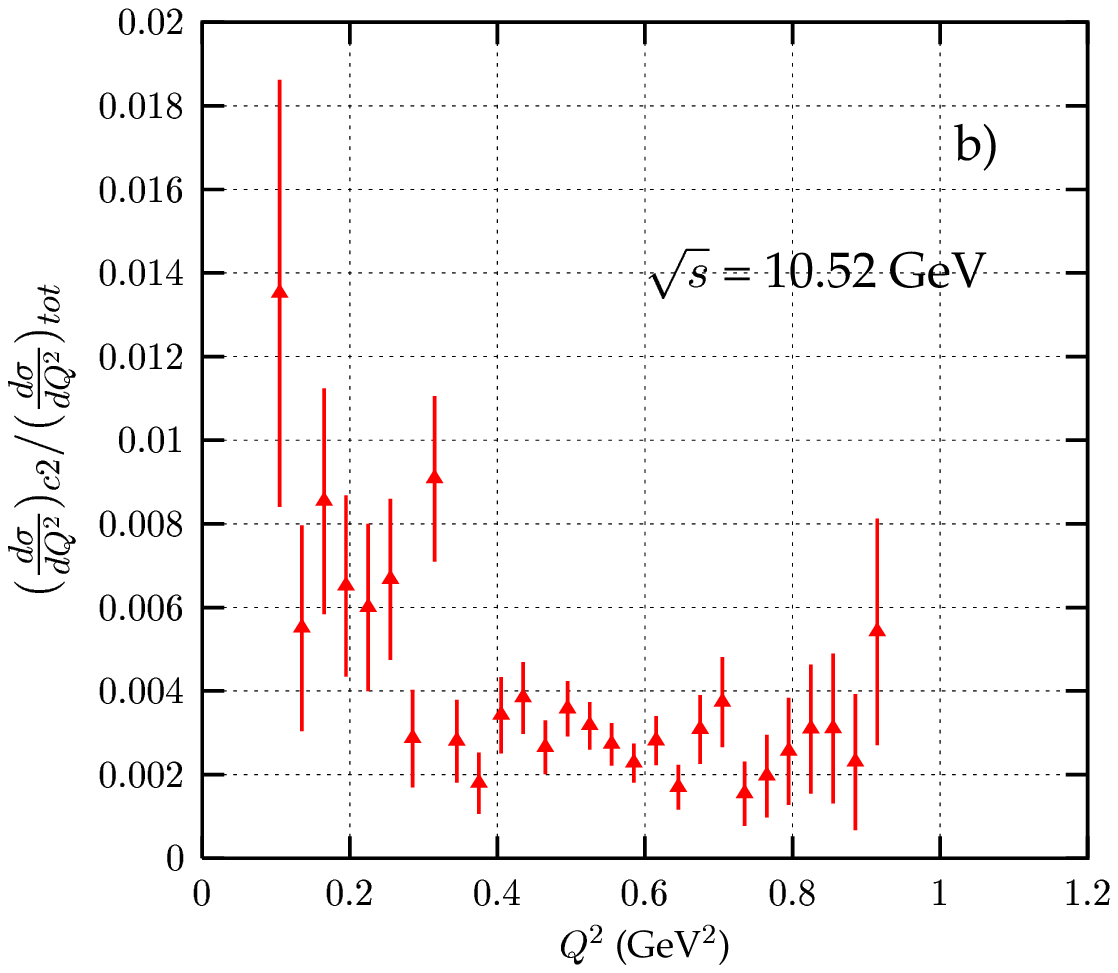,width=6.4cm,height=5.8cm}
\end{center}
\caption{
Non-reducible pair production background 
 at $B$--factory energy : (a) no cuts in $ ( \frac{d\sigma}{dQ^2})_{tot}$;
 (b) for both $ (\frac{d\sigma}{dQ^2})_{c2} $ and 
  $ ( \frac{d\sigma}{dQ^2})_{tot}$ cuts on pion angles are imposed:
   $  50^\circ < \theta_{\pi^{\pm}} < 130^\circ $.
 For (a) and (b) c2 denotes additional cuts:
 $ (\theta_{e^+} < 20^\circ $ or $  \theta_{e^+} > 160^\circ) $
and $ (  \theta_{e^-} < 20^\circ $ or $  \theta_{e^-} > 160^\circ)  $.
}
\label{f5}
\end{figure}

 The situation is completely different at $B$--factories, where one can
 almost completely reduce the background coming from $e^+e^-$ pairs
 ( Fig.\ref{f5}), by rejecting the events with at least one charged
 lepton, electron or positron, in the detector.

\section{Conclusions}
 We have constructed the Monte Carlo generator EKHARA,
 which simulates the 
 reaction $e^+e^- \to \pi^+\pi^- e^+e^-$ 
 based on initial and final state emission of a $e^+e^-$ pair
 from $e^+e^-\to\pi^+\pi^-$ production diagram.
 A detailed study of this process, as a potential background
 for $\sigma (e^+e^- \to \pi^+\pi^-)$ measurement
 via radiative return method,
 shows that it can become important, when 
 the experimental precision will reach 1\%, or better.

 {\bf Acknowledgements:}
 We thank J.H. K{\"u}hn for discussion and
 we are grateful for the support and the kind hospitality of
 the Institut f{\"u}r Theoretische Teilchenphysik 
 of the Universit\"at Karlsruhe.

\end{document}